\documentclass[a4paper,12pt]{article}
\usepackage{amsmath}
\usepackage{amssymb}
\usepackage{braket}
\usepackage{cancel}
\usepackage{graphicx}
\usepackage{color}
\usepackage[utf8]{inputenc}
\usepackage{booktabs}
\usepackage{hyperref}
\usepackage{multirow}
\usepackage{slashed}
\usepackage[normalem]{ulem}
\usepackage{wasysym}
\usepackage{amssymb}
\usepackage{braket}
\usepackage{simplewick}
\usepackage{array}
\usepackage{graphicx}
\usepackage{float}
\usepackage{titlesec}
\usepackage{listings}
\usepackage{hyperref}
\usepackage[english]{babel}
\usepackage{xcolor}
\usepackage{xspace}
\usepackage{subcaption}

\newcommand{\muG}{\ensuremath{\mu_{G}^2}\xspace}
\newcommand{\mupi}{\ensuremath{\mu_\pi^2}\xspace}

\newcommand{\rhoD}{\ensuremath{\rho_{D}^3}\xspace}
\newcommand{\rhols}{\ensuremath{\rho_{LS}^3}\xspace}

\usepackage[english]{babel}

\usepackage{graphicx}

\usepackage{geometry}
\usepackage{romannum}

\renewcommand{\arraystretch}{1.25}

\begin{document}
\pagenumbering{arabic}

\begin{titlepage}

\begin{flushright}
{\small
P3H-22-069  \\ 
SI-HEP-2022-14  \\
\today \\
}
\end{flushright}

\vskip1cm
\begin{center}
{\Large \bf\boldmath 
Standard Model predictions for Lepton Flavour Universality ratios of inclusive semileptonic $B$ decays}
\end{center}

\vspace{0.5cm}
\begin{center}
{Muslem Rahimi$^a$ and K. Keri Vos$^{b,c}$} \\[6mm]

{

{\it $^a$ Center for Particle Physics Siegen (CPPS), \\
Theoretische Physik 1, Universit{\"a}t Siegen, \\
57068 Siegen, Germany}\\[0.3cm]

{\it $^b$Gravitational 
Waves and Fundamental Physics (GWFP),\\ 
Maastricht University, Duboisdomein 30,\\ 
NL-6229 GT Maastricht, the
Netherlands}\\[0.3cm]

{\it $^c$Nikhef, Science Park 105,\\ 
NL-1098 XG Amsterdam, the Netherlands}}
\end{center}

\vspace{0.6cm}
\begin{abstract}
\vskip0.2cm\noindent
We present Standard Model predictions for lepton flavour universality ratios of inclusive $B\to X_{(c)} \ell \bar\nu_\ell$. For the $\ell=\mu,e$, these ratios are very close to unity as expected. For the $\tau$ mode, we update the SM prediction for the branching ratio including power-corrections in the heavy-quark expansion up to $1/m_b^3$. These inclusive ratios serve as an important cross-check of the exclusive $B\to D^{(*)}\ell\bar\nu_\ell$ modes, in which tensions exists between the predictions and measurements in those modes. 
\end{abstract}

\end{titlepage}

\section{Introduction}
\label{sec::Introduction}
The inclusive $B\to X_c \ell \bar\nu_\ell$ decays, with $\ell = \mu,e$, are by now standard candles in the determination of the CKM element $|V_{cb}|$. Employing the heavy quark expansion (HQE), allows the parametrization of these decays in perturbative Wilson coefficients and non-perturbative HQE elements. Thanks to a combined theoretical and experimental effort, these HQE parameters can be extracted from moments of the decay spectrum giving an impressive $2\%$ uncertainty on the inclusive $V_{cb}$ determinations \cite{Bordone:2021oof, Bernlochner:2022ucr}.

The experimental measurements of semileptonic $B\to X_c$ usually combine the muon and electron modes (and $B^0$ and $B^+$). Recently, the Belle collaboration also provided the first measurement of $q^2$ moments, separately for the electron and muon modes \cite{Belle:2021idw}. No deviations from lepton flavor universality were found. However, given the discrepancies in the rare $b\to s\ell\ell$ modes, it may be worth measuring the ratio
\begin{equation}
\label{eq:ratdef}
R_{\mu/e}(X_c) \equiv \frac{\Gamma(B\to X_c \mu \bar\nu_\mu)}{\Gamma(B\to X_c e \bar\nu_e)} \ .
\end{equation}
In the Standard Model (SM), this ratio is expected to be close to one, but more elaborate predictions are not available to our knowledge. In this paper, we provide these predictions by taking into account the masses of the leptons, in light of upcoming measurements. We do not include structure depend or ultrasoft QED effects as those are challenging to disentangle from the experimental detector efficiencies (for recent works on QED effects in exclusive semileptonic $B$ decays see e.g. \cite{Beneke:2020vnb, Beneke:2021jhp, Papucci:2021ztr, Cali:2019nwp}.). We leave a detailed discussion of the QED effects in inclusive decays for future works. 

While the light-lepton modes have been studied in depth, the situation is very different for the $\tau$ mode. Experimentally, only LEP results \cite{pdg} and a unpublished Belle analysis \cite{Hasenbusch:2018pxj} of the total rate exists, both having large uncertainties. In addition, the LEP measurement requires assumptions about hadronic effects in order to be interpreted. On the theoretical side, SM predictions for this mode exists using the HQE parameters as input. In this paper, we update these predictions to include HQE parameters up to $1/m_b^3$, which have a relatively large impact. These higher-order terms were first studied in \cite{Mannel:2017jfk}, but this reference misses some terms in the $\rho_D^3$ coefficient. Here we correct these results. We point out that numerically, the difference between our results and \cite{Mannel:2017jfk} is small. In light of the tensions in ratios of the exclusive $B \to D^{(*)} \ell \bar\nu_\ell$ versus $B\to D^{(*)} \tau \bar\nu_\tau$ (see e.g.~\cite{Bernlochner:2022ywh} for a recent review on semileptonic $\tau$ modes), we stress the importance of an independent cross-check in the inclusive channel. For this, the SM predictions derived in this short letter are vital. These predictions can be used in the search for new physics, especially in the tau sector where new measurements are expected soon.

\section{Inclusive decay of $b \to c \ell \bar\nu_\ell$ with massive leptons}
\label{sec::threebodydecay}
To calculate the inclusive $b \to c$ semileptonic rate, we employ the standard heavy-quark expansion (HQE). This allows us to perform an operator product expansion (OPE) for the triple differential rate in the lepton (neutrino) energy $E_{\ell (\nu)}$ and the dilepton invariant mass $q^2$ as
\begin{equation}
    \frac{d\Gamma}{dE_\ell dq^2 dE_\nu} = \frac{G_F^2 |V_{cb}|^2}{16 \pi^3} L_{\mu\nu}W^{\mu\nu} \ .
\end{equation}
Here $L_{\mu\nu}$ is the lepton tensor and $W^{\mu\nu}$ the hadronic tensor as defined in e.g.~\cite{Mannel:2021mwe}. Expressing the $W^\mu\nu$ tensor in Lorentz scalars as usual then gives
\begin{align}
\label{eq:tripdif}
    \frac{\text{d}\Gamma}{\text{d}E_{\ell} \text{d}q^2 \text{d}E_{\nu}} &= \frac{G_{F}^2 |V_{cb}|^2}{2 \pi^3} \left[q^2 W_{1} + (2 E_{\ell} E_\nu - \frac{q^2}{2}) W_2 + q^2 (E_{\ell} - E_{\nu}) W_3 \right. \nonumber \\
    & \left. \frac{1}{2} m_{\ell}^2 \left( -2 W_{1} + W_2 -2 (E_\nu+ E_\ell)W_{3} + q^2 W_4 + 4 E_\nu W_5 \right) - \frac{1}{2} m_{\ell}^4 W_4 \right] \ ,
\end{align}
where we have omitted explicit $\theta$-functions (see \cite{Ligeti:2014kia}). 

In general, for $B\to X_c \mu \bar\nu_\mu$ and $B\to X_c e \bar\nu_e$, lepton masses are neglected. However, for the much heavier decay involving the $\tau$ lepton: $B\to X_c \tau \bar\nu_\tau$, such an approximation cannot be made. We calculated the total inclusive rate including lepton masses. This calculation differs from the standard case, as now also the structure functions $W_4$ and $W_5$ in \eqref{eq:tripdif} contribute and because the phase space boundaries are affected. We refer to \cite{Mannel:2017jfk, Ligeti:2014kia} for details. 

Considering terms up to $1/m_b^3$, we write the total rate as
\begin{equation}\label{eq:rate}
\Gamma(B \to X_c \ell\bar \nu_\ell) =  \Gamma_0\!\!
\left[C_0^{(0)} + \frac{\alpha_s}{\pi} C_0^{(1)}
+ C^\perp_{\mu_\pi^2} \cfrac{(\mu_\pi^2)^\perp}{m_b^2} + C^\perp_{\mu_G^2} \cfrac{(\mu_G^2)^\perp}{m_b^2}
+ C^\perp_{\rho_D^3}\cfrac{(\rho_D^3)^\perp}{m_b^3} + C^\perp_{\rho_{LS}^3}\cfrac{(\rho_{LS}^3)^\perp}{m_b^3} \right], 
\end{equation}
where the coefficients depend on 
\begin{equation}
    \rho \equiv m_c^2/m_b^2 \ , \quad\quad\quad \eta \equiv m_\ell^2/m_b^2 \ ,
\end{equation}
and
%
\begin{equation}
    \Gamma_0 \equiv \frac{G_F^2|V_{cb}|^2 m_b^5}{192\pi^3}\, (1+A_{\text{ew}}), 
\end{equation}
which includes the electroweak correction $A_{\rm ew}=0.014$\cite{Sirlin:1974ni}. 

We define the nonperturbative parameters as (see e.g.~\cite{Mannel:2010wj})
\begin{eqnarray}
2 m_B \, (\mu_\pi^2)^\perp & \equiv & 
- \langle B |\bar b_v (iD_\rho) (iD_\sigma) b_v| B \rangle \Pi^{\rho\sigma} \ ,
\\
2 m_B \, (\mu_G^2)^\perp & \equiv & \frac{1}{2}
\langle B |\bar b_v \left[i D_\rho, i D_\lambda\right] (- i \sigma_{\alpha \beta}) b_v| B \rangle  \Pi^{\alpha \rho}  \Pi^{\beta\lambda},
\\
2 m_B \, (\rho_D^3)^\perp & \equiv & \frac{1}{2}
\langle B |\bar b_v \left[i D_\rho, \left[iD_\sigma, iD_\lambda\right]\right] b_v| B \rangle \Pi^{\rho\lambda}  v^\sigma,
\\
2 m_B \, (\rho_{LS}^3)^\perp & \equiv & \frac{1}{2}
\langle B |\bar b_v\left\{i D_\rho, \left[iD_\sigma, iD_\lambda\right]\right\}(-i\sigma_{\alpha\beta}) b_v| B \rangle \Pi^{\alpha\rho}\Pi^{\beta\lambda}  v^\sigma\, ,
\end{eqnarray}
where
\begin{equation}
    \Pi_{\mu\nu} = g_{\mu\nu}-v_{\mu}v_\nu \ . 
\end{equation}
The above definitions differ from e.g.~\cite{Mannel:2010wj} and \cite{Mannel:2018mqv, Fael:2018vsp} where the full covariant derivative was used and not only the spatial component as above, linked via $iD_\mu= v_\mu ivD + D_\perp$. To differentiate, we therefore add a $\perp$ superscript to HQE parameters.  
The relation between the ``perped'' and full covariant derivative parameters is
\begin{equation}\label{eq:mugperp}
  (\mu_G^2)^\perp = \mu_G^2 + \frac{\rhoD + \rhols}{m_b} \ ,
\end{equation}
while $(\mupi)^\perp = \mupi, (\rhols)^\perp = \rhols$ and $(\rhoD)^\perp = \rhoD$ up to terms of order $1/m_b^3$ (see discussion in Appendix~A of \cite{Fael:2018vsp}). 

We list all coefficients, except $C_0^{(1)}$ in Appendix~A, for completeness. Setting $\eta\to 0$, reproduces the well-known rate  \cite{ Mannel:2018mqv,Gremm:1996df, Dassinger:2006md}

The coefficients agree with \cite{Mannel:2017jfk} (and previous results in \cite{Falk:1994gw,Balk:1993sz} for $C_0, C_{\mupi}$ and $C_{\muG}$) up to a difference in the $C_{\rhoD}$. The discrepancy with \cite{Mannel:2017jfk} arises due to the more involved integrations which now contain additional delta functions. For the total rate, where no cut on lepton energy is required, it is easiest to first perform the integration over the lepton energy $E_\ell$ analytically (as the structure functions $W$ do not depend on $E_\ell$.). In the limit $\rho=\eta$, our calculation can be checked and agrees with \cite{King:2021xqp}. We have also contacted the authors of \cite{Mannel:2017jfk}, who now agree with our results.  

We recalculated the perturbative corrections for the partonic rate $C_0^{(1)}$ which agree with  \cite{Czarnecki:1994bn,Jezabek:1996db}.  Our analysis does not include $\alpha_s^2$ corrections, which are known~\cite{Biswas:2009rb} but only available for fixed $m_b/m_c$. To fully include such effects in a state-of-the-art manner, a new analysis is required. We briefly discuss these corrections in the following. We note that for $\eta=0$, these corrections are even known up to $\alpha_s^3$ \cite{Fael:2022frj}. 

\section{SM predictions for inclusive rates including masses}
With the coefficients $C_i$ for the total rate, we can now in principle predict the branching ratios for semileptonic $b\to c$ decays. However, the light lepton decays and their moments are used to determined the HQE parameters and $V_{cb}$. Therefore, such predictions are not very instructive for light mesons. For those, we therefore restrict ourselves to ratios of semileptonic modes. For the tau modes, we also discuss the total branching ratio. 

For our numerical analysis we use the input values listed in Table~\ref{tab:input} obtained from  \cite{Bordone:2021oof}. As is customary, we work in the kinetic mass scheme, which can be related to the pole mass via a perturbative series~\cite{Gambino:2004qm,Gambino:2007rp,Fael:2020iea}.

\begin{table}[t]
\begin{center}
\begin{tabular}{ c | c }
\hline\hline
$m^{\text{kin}}_{b}$ & (4.573 $\pm$ 0.012) GeV 
\\
\hline
$\overline{m}_{c}$(2 GeV) & (1.092 $\pm$ 0.008) GeV  
\\ 
\hline
$\left(\mu_{\pi}^{2}(\mu)\right)_{\text{kin}}$ & (0.477 $\pm$ 0.056) GeV$^{2}$
\\
\hline
$\left(\mu_{G}^{2}(\mu)\right)_{\text{kin}}$ & (0.306 $\pm$ 0.050) GeV$^{2}$ 
\\
\hline
$\left(\rho_{D}^{3}(\mu)\right)_{\text{kin}}$ & (0.185 $\pm$ 0.031) GeV$^{3}$
\\
\hline
$\left(\rho_{LS}^{3}(\mu)\right)_{\text{kin}}$ & (-0.130 $\pm$ 0.092) GeV$^{3}$
\\
\hline
$V_{cb}$ & $ (42.16 \pm 0.51 ) \cdot 10^{-3}$ \\
\hline\hline
\end{tabular}
\end{center}
\caption{Numerical inputs taken from \cite{Bordone:2021oof}, where the HQE parameters are defined in the perp basis. For the charm mass, we use the $\overline{\text{MS}}$ scheme at 2 GeV. All other hadronic parameters are in the kinetic scheme at $\mu = 1$ GeV.}
\label{tab:input}
\end{table}

\subsection{Lepton Flavour Universality Ratios}
We define the ratios $R_{\mu/e}$ as in \eqref{eq:ratdef} and define equivalently $R_{\tau/\mu}$ and $R_{\tau/e}$. In such ratios, $V_{cb}$ drops out, but the HQE parameters do not completely, due to different mass effects. Splitting the contributions to $R(X_c)$ according to
\begin{align}
R(X_c) &= \xi_{\text{LO}} + \xi_{\text{NLO}} \left(\frac{\alpha_s}{\pi} \right) +\xi_{\mu_G^2} \,(\mu_G^2)^\perp + \xi_{\mu_\pi^2} \, (\mu_\pi^2)^\perp + \xi_{\rho_{LS}^3} \, (\rho_{LS}^3)^\perp + \xi_{\rho_{D}^3} \, (\rho_{D}^3)^\perp  \ ,
 \label{eq::observables}
\end{align}
we find the SM predictions listed in Table~\ref{tab:observable}. The uncertainties in Table \ref{tab:observable} are obtained by combining all uncertainties of the input parameters in quadrature. In addition, we vary the scale of $\alpha_s(\mu)$ from $m_b/2< \mu < 2m_b$.  We note that $\mu_\pi^2$ completely drops out in such ratios, while the effect of $\rhoD$ is relatively large even though this is a $1/m_b^3$ contribution. We do not include an additional uncertainty for missed higher-order terms of order $1/m_b^4$ and beyond.

\begin{table}[t]
\begin{center}
\renewcommand{\arraystretch}{1.2} 
\begin{tabular}{c c c c }
\toprule
 & $R_{\tau/\mu}(X_c) \cdot 10^{-2}$ & $R_{\tau/e}(X_c) \cdot 10^{-2}$ & $R_{\mu/e}(X_c) \cdot 10^{-2}$ \\
\midrule
$\xi_{\text{LO}}$ & 23.557 & 23.429 & 99.458   \\
\hline
$\xi_{\text{NLO}}$ & 5.446 & 5.451 & 0.144   \\
\hline
$\xi_{\mu_G^2}$ & -2.165 & -2.161 & -0.0315 \\
\hline
$\xi_{\mu_\pi^2}$ & 0 & 0 & 0   \\
\hline
$\xi_{\rho_{LS}^3}$ & 0.4735 & 0.4726 & 0.0068   \\
\hline
$\xi_{\rho_{D}^3}$ & -6.785 & -6.765 & -0.0709   \\
\hline
& $21.965 \pm 0.420 $ & $21.843 \pm 0.419$ & $99.445 \pm 0.006$ \\
\toprule
\end{tabular}
\caption{SM predictions for the inclusive LFU ratios. We list the different contributions separately according to (\ref{eq::observables}). The uncertainty is obtained by varying all input parameters and adding those in quadrature. }
\label{tab:observable}
\end{center}
\end{table}

For the $\tau$ modes, we find
\begin{align}\label{eq:RXctau}
    R_{\tau/\mu}(X_c)|_{\text{NLO} + 1/m_b^2 + 1/m_b^3} &= 0.220 \pm 0.004 \nonumber \\
     R_{\tau/e}(X_c)|_{\text{NLO} + 1/m_b^2 + 1/m_b^3} &= 0.218 \pm 0.004 \ .
\end{align}
This is in agreement with previous determination in \cite{Freytsis:2015qca}, which includes terms up to $1/m_b^2$ in the 1S-scheme:
\begin{equation}\label{eq:RXcZoltan}
    R(X_c)_{\rm FLR} = 0.223 \pm 0.004 \ .
\end{equation}
In this case, the uncertainty is dominated by $m_b$ and $\lambda_1$ (i.e. the HQE element in the infinite mass limit) and includes an additional uncertainty of half of the $\alpha_s^2$ term. It also does not include an additional uncertainty due the missed $1/m_b^3$ terms. 

Finally, also a calculation of only the partonic rates at  $\mathcal{O}(\alpha_s^2)$ exists ~\cite{Biswas:2009rb} 
\begin{align}
    R(X_c)_{\text{BM}} &= 0.237 \pm 0.031 \ ,
    \label{eq:RXc-BM}
\end{align}
which is based on the on-shell scheme. It was found that $\alpha_s^2$ effects in the $R_{\tau/\ell}(X_c)$ ratio are very small. While the ratio of leading order decay rates is a rapidly changing function of $m_b, m_c$ and $m_\tau$, radiative corrections to $\mathcal{B}( B \to X_c \tau \nu) $ and $\mathcal{B}(B \to X_c \ell \nu) $ are correlated, so they cancel out in the ratio that is largely independent of the quark masses. Here we do not include these $\alpha_s^2$ effects as \cite{Biswas:2009rb} only provides them at fixed $m_c/m_b$. However, we have verified that the $\alpha_s^2$ corrections are only 2$-$3 \% of the NLO order contribution. Therefore, our uncertainty estimate obtained by varying $\alpha_s$ accounts for these effects. We also note that our $\alpha_s$ corrections are half of those in \cite{Biswas:2009rb}, due to the switch to the kinetic scheme.

\subsection{Ratios for semileptonic $B\to X$}
\label{sec:BtoX}
Experimentally, in order to obtain the semileptonic $B\to X_c$, the $B\to X_u$ background has to be dealt with. On the other hand, as pointed out in \cite{Mannel:2021mwe}, this $V_{ub}^2/V_{cb}^2$ suppressed contribution can also be calculated in the local OPE. Naively taking the $B\to X_c$ rate and setting $\rho\to 0$ works up to $1/m_b^2$, but at order $1/m_b^3$ additional four-quark operators (weak annihilation) have to be introduced that cure the divergence arising in the $\rhoD$ term (see e.g.~\cite{Fael:2019umf} for references and discussions). For charm, such effects were studied in \cite{Gambino:2010jz} using semileptonic $D$ meson data from CLEO \cite{CLEO:2009uah}. For $B\to X_u$, this issue will be discussed specifically in an upcoming publication \cite{WorkInProgress3}. However, at the moment, we can make a reliable estimate for the $R(X)$ ratio by calculating the $B\to X_u$ effects by setting $\rhoD\to 0$. We then have
\begin{equation}
    \Gamma(B\to X \ell \bar\nu_\ell ) = \Gamma(B\to X_c \ell \bar\nu_\ell ) + \left( \frac{|V_{ub}|}{|V_{cb}|}\right)^2 \Gamma(B\to X_c \ell \bar\nu_\ell )|_{\rho\to 0, \rhoD\to 0} \ .
\end{equation}
To derive ratios of the $B\to X$ semileptonic rates, we use the exclusive $V_{ub}$ determination from \cite{Leljak:2021vte}:
\begin{equation}
    V_{ub}|_{\rm excl.} = (3.77 \pm 0.15) \cdot 10^{-3} \ ,
\end{equation}
which is in agreement at the $1-2\sigma$ level with the recent inclusive determinations \cite{Belle:2021eni}. For $V_{cb}$, we take the recent inclusive determination in $V_{cb} = (42.16 \pm 0.51)\cdot 10^{-3}$ \cite{Bordone:2021oof}. 

We then find
\begin{align}\label{eq:RXstuff}
     R_{\tau/\mu}(X) &= 0.221 \pm 0.004 \, ,\\
    R_{\tau/e}(X) &= 0.220 \pm 0.004 \, ,\\
    R_{\mu/e}(X) &= 0.994 \pm 0.001 \, .
\end{align}
We do not quote the $R(X_u)$ as there we do not have the $V_{ub}^2$ suppression. As such, weak annihilation and $\rhoD$ effects may play a bigger role. 

Finally, we note that experimentally, usually a lower cut on the lepton energy $E_\ell$ employed. Alternatively, also a $q^2$ cut can be imposed, as suggested first in \cite{Fael:2018vsp}, where $q^2$ moments of the spectrum are advertised. A $q^2$ cut is easier to implement for the $\alpha_s$ corrections, therefore we also quote ratios with such a cut. Here we take $q^2_{\rm cut}=3$ GeV$^2$ as a default cut. The full expression with an arbitrary $q^2_{\rm cut}$ can be provide by the authors. 
We find
\begin{align}
    R_{\tau/\mu}(X)_{q^2_{\rm cut}} &= 0.352 \pm 0.004 \, ,\\
    R_{\tau/e}(X)_{q^2_{\rm cut}} &= 0.352 \pm 0.004 \, ,\\
    R_{\mu/e}(X)_{q^2_{\rm cut}} &= 0.999 \pm 0.001 \, ,
\end{align}



\subsection{Inclusive decay of $b\to c\tau\bar\nu_\tau$}

\begin{table}[t]
\begin{center}
\renewcommand{\arraystretch}{1.2} 
\begin{tabular}{c c c}
\toprule
 &  $\mathcal{B}(B \to X_c \tau \bar\nu_\tau)$ $[\%]$ & $\mathcal{B}(B \to X \tau \bar\nu_\tau)$ $[\%]$ \\
\midrule
$\xi_{\text{LO}}$ & 3.042  & 3.095\\
\hline
$\xi_{\text{NLO}}$ &  -3.064  & -3.020 \\
\hline
$\xi_{\mu_G^2}$ & -0.557 & -0.564 \\
\hline
$\xi_{\mu_\pi^2}$ & -0.0727 & -0.074\\
\hline
$\xi_{\rho_{LS}^3}$ & 0.122 & 0.123  \\
\hline
$\xi_{\rho_{D}^3}$ &  -1.408  & -1.408\\
\hline
&  $2.341 \pm 0.130$ & $2.395 \pm 0.131$ \\
\toprule
\end{tabular}
\caption{Predictions for the branching ratio within the local OPE, using  $V_{cb} = (42.16 \pm 0.51 ) \cdot 10^{-3}$ \cite{Bordone:2021oof}. We quote the flavour-averaged rate. Predictions for the charged or neutral $B$ decay can be obtained by multiplying with $\tau_{B^{+,0}}/\tau_{B}$. }
\label{tab:brtau}
\end{center}
\end{table}
Using \eqref{eq:rate}, we update the SM predictions for the $\tau$-mode. Taking the recent $V_{cb} = (42.16 \pm 0.51)\cdot 10^{-3}$ \cite{Bordone:2021oof}, and the HQE inputs in Table~\ref{tab:input}, we find the contributions given in Table~\ref{tab:brtau}. These results use the averaged decay rate $\tau_{B} = 1.579 $ ps \cite{pdg}, which can be adjusted for the $B^{+,0}$ by multiplying with $\tau_{B^{+,0}}/\tau_{B}$. In addition, predictions for the recent determination of $V_{cb}$ from $q^2$ moments: $V_{cb}= (41.69\pm 0.63)\cdot 10^{-3}$ can be obtained by rescaling.

Calculating the branching ratio directly from the OPE gives
\begin{align}
\label{eq:brdir}
 \mathcal{B}&(B\to X_c \tau\bar\nu_\tau)_{\rm OPE} = \nonumber \\
 &\left( 2.34 \pm 0.07|_{m_b} \pm 0.03|_{m_c} \pm 0.02|_{\mu_G^2} + 0.01|_{\rho_{LS}^3} +0.04|_{\rho_D^3} + 0.06|_{\alpha_s} + 0.05|_{V_{cb}}\right) \% \nonumber \\
 &=(2.34 \pm 0.13)\%\ \, ,
 \end{align}
 where we specify the different contributions to the uncertainty and in the last line we summed these in quadrature. Again, we do not include an additional uncertainty due to missed higher-order terms. 
For completeness we also quote the $B^+$ and $B^0$ rates separately
\begin{align}
\mathcal{B}(B^+\to X_c^+ \tau\bar\nu_\tau)= (2.43 \pm 0.13)\%\ . \nonumber \\
\mathcal{B}(B^0\to X_c^0 \tau\bar\nu_\tau)= (2.25 \pm 0.13)\%\ .
\end{align} 
Our value agrees with \cite{Mannel:2017jfk}, despite a missed $\rho_D^3$ contribution in that paper. Finally, following the discussion in Sec.~\ref{sec:BtoX}, we find the $B\to X$ rate as
\begin{align}\label{eq:opeX}
    \mathcal{B}(B \to X \tau \nu) &=  \left( 2.39 \pm 0.13 \right) \% \, .
\end{align}

These determinations are in agreement with the LEP measurement of the inclusive branching fraction of the admixture of bottom baryons \cite{pdg}
\begin{equation}
{\mathcal{B}} (b\mbox{-admix}  \to X \tau \bar\nu_\tau) = (2.41 \, \pm 0.23) \% \, ,
\label{eq:LEP-data}
\end{equation} 
which only to leading order in the HQE can be interpreted as the individual hadron rates. 

In addition, there exists an unpublished Belle measurement of the $R_{\tau/(e,\mu)}(X)$ \cite{Hasenbusch:2018pxj}:
\begin{align}
R(X) &\equiv \frac{\mathcal{B}(B \to X \tau \bar\nu_\tau)}{\mathcal{B}(B \to X \ell \bar\nu_\ell)} = 0.298 \pm 0.022 \, ,
\label{eq:belle}
\end{align} 
where $\ell = \mu,e$. Comparing this with our estimate in \eqref{eq:RXstuff}, we observe a slight tension. Alternatively, we may also estimate the relation between $R(X)$ and $R(X_c)$, by subtracting the theoretically calculated rate. We find 

\begin{align}
    R(X) &= 
     \begin{cases}
       R_{\tau/\mu}(X_c) \left(1+ {\rm 1.012} \dfrac{|V_{ub}|^2}{|V_{cb}|^2}\right)& \text{for $\ell = \mu$  ,}\\
      R_{\tau/e}(X_c) \left(1+ {\rm 1.014} \dfrac{|V_{ub}|^2}{|V_{cb}|^2}\right)& \text{for $\ell = e$  .}
    \end{cases}      
\end{align}
Therefore, we will interpret $R(X) = R(X_c)$. Comparing then \eqref{eq:belle} with our predictions in \eqref{eq:RXctau}, we again observe a slight tension. 

Besides calculating the rate directly from the OPE as in \eqref{eq:brdir}, we may also give predictions of the branching ratio by multiplying them with the measured flavor-averaged light-meson branching ratio. Following the detailed discussion in \cite{Bernlochner:2022ywh}, we take 
\begin{equation}
    \mathcal{B}(B\to X_c \ell \bar\nu_\ell) = (10.48 \pm 0.13)\% ,
    \label{eq:brmeas}
\end{equation}
which differs slightly from those quoted by  \cite{Bordone:2021oof} and \cite{pdg}. 
Averaging our predictions for the muon and electron ratios in \eqref{eq:RXctau}, and multiplying with \eqref{eq:brmeas}, we find 
\begin{align}
\mathcal{B} (B \to X_c \tau \bar\nu_\tau)_{\text{Exp+OPE}} &\equiv \mathcal{B}(B\to X_c \ell \bar\nu_\ell)  \, R_{\tau/\ell}(X_c) = (2.30 \pm 0.05) \% \, ,
\label{eq:brfromRc}
\end{align}
which is in perfect agreement with, but has a much smaller uncertainty than our direct calculation in Eq. \eqref{eq:brdir}. 

Similarly, we can convert the unpublished Belle measurement in 
Eq. \eqref{eq:belle}. In \cite{Hasenbusch:2018pxj}, this is multiplied with the measured isospin-average branching fraction $\mathcal{B} (B \to X \ell \bar\nu_\ell) = (10.86 \pm 0.16)\%$ to obtain $\mathcal{B} (B\to X \tau \bar\nu_\tau) = (3.23 \pm 0.25)\% $. This is in tension with the value we find from the direct OPE calculation in Eq. \eqref{eq:opeX}. Using Eq. \eqref{eq:belle}, we multiply with Eq. \eqref{eq:brmeas} to find 
\begin{align}
\mathcal{B} (B\to X_c \tau \bar\nu_\tau)_{\rm Belle} &= (3.12\pm 0.23)  \, \% 
\end{align}

Similarly, we can convert the previous theoretical determination of $R(X_c)$ in Eq. \eqref{eq:RXcZoltan} \cite{Freytsis:2015qca} with this rate. We find
\begin{equation}
    \mathcal{B}(B \to X_c \tau \bar\nu_\tau)_{\rm FLR} = (2.34 \pm 0.05)\% , 
\end{equation}
which is in agreement with the value reported in \cite{Bernlochner:2021vlv}.


Multiplying the branching ratio in Eq. (\ref{eq:brmeas}) with Eq. (\ref{eq:RXc-BM}) we obtain:
\begin{align}
     \mathcal{B}(B \to X_c \tau \bar\nu_\tau)_{\rm BM} = (2.47 \pm 0.04)\% , 
\end{align}

\begin{figure}[t]
	\centering
	\subfloat{\includegraphics[width=0.8\textwidth]{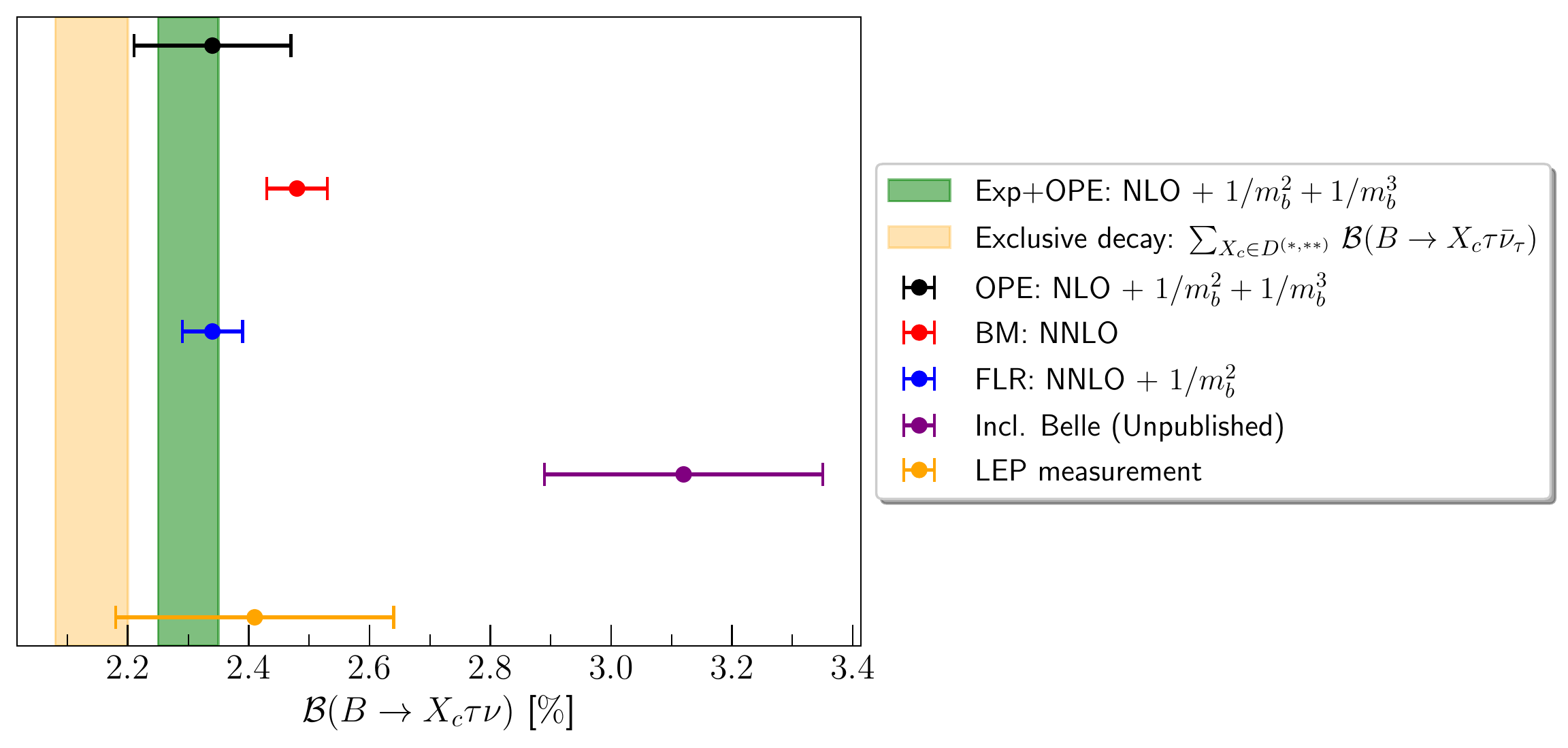}}
  \caption{Comparison of our predictions for the branching ratio $\mathcal{B}(B \to X_c \tau \nu)$ with previous determinations and with the sum over exclusives from \cite{Bernlochner:2021vlv}. We also quote the measurements of LEP and the unpublished Belle measurement (see text for details).  }
  \label{fig:sum}
\end{figure}

Finally, it is also interesting to compare our inclusive predictions with a sum over exclusive. To this extend, we follow the recent \cite{Bernlochner:2021vlv}. Using the HFLAV-averaged SM predictions for $R(D)$ and $R(D^*)$ and the measured rates for the light-modes, combined with the prediction for $\mathcal{B}(B\to D^{**} \ell \bar\nu_\ell)$  \cite{Bernlochner:2017jka}, they find \cite{Bernlochner:2021vlv}
\begin{equation}
    \sum_{X_c \in D^{(*,**)}} \mathcal{B}(B \to X_c \tau \bar\nu_\tau) = (2.14\pm 0.06)\; \% \ .
\end{equation}
Interestingly, this sum over exclusive modes does not saturate our calculated fully inclusive rate. We summarize and visualize our findings in Fig.~\ref{fig:sum}.

\section{Conclusion}
\label{sec::Conclusion}
We calculated the SM predictions for the lepton flavour universality ratios of semileptonic inclusive $B$ decays. In these predictions, we only considered the mass effects, and included HQE parameters up to $1/m_b^3$. We corrected a previous calculation in \cite{Mannel:2017jfk}, which missed some terms in the $\rhoD$ contribution. 

In addition, we present updated results of the Standard Model for the branching ratio of the $B \to X_c \tau \bar\nu_\tau$ decay. Experimentally, for this rate only a LEP measurement and an unpublished Belle analysis are available. In light of the discrepancies between data and experiment in the universality ratios of exclusive semileptonic $B\to D^{(*)}$ update measurements of this observable are highly wanted. A detailed analysis of the effect of new physics operators on inclusive semitauonic decays is in progress \cite{WorkInProgress}.

\subsubsection*{Acknowledgements}
We thank Florian Bernlochner for suggesting this project in light of upcoming measurements. Additionally, we thank him, Thomas Mannel and Matteo Fael for their comments. We thank A. Rusov for discussions about \cite{Mannel:2017jfk}. This  research  was supported by the Deutsche Forschungsgemeinschaft (DFG, German Research Foundation) under grant 396021762 - TRR 257.

\section{Appendix}
In this Appendix, we explicitly give the coefficients of the rate in \eqref{eq:rate}. 

We note that all these coefficients except $C_{\rho_D^3}$ agree with \cite{Mannel:2017jfk} when transforming basis from the spatial derivative ``perped'' basis used here to the full covariant derivative basis via \eqref{eq:mugperp}. Explicitely this means that
\begin{equation}
    C_{\mu_\pi^2} = C_{\mu_\pi^2}^\perp \ , C_{\mu_G^2} = C_{\mu_G^2}^\perp \ , C_{\rho_D^3} = C_{\rho_D^3}^\perp + C_{\mu_G^2}^\perp \ , C_{\rhols} = 0 \ .    
\end{equation}

We find
\begin{eqnarray}
C_0^{(0)} & = &
R \left[1 - 7 \rho - 7 \rho^2 + \rho^3 - (7 - 12 \rho + 7 \rho^2) \eta  
- 7 (1 + \rho) \eta^2 + \eta^3 \right] \\
& - & 
12 \left[\rho^2\,
  \ln \frac{(1 + \rho - \eta - R)^2}{4\rho} 
- \eta^2\,\text{ln}\frac{(1+\eta-\rho +R)^2}{4\eta}
- \rho^2 \eta^2 \, \ln \frac{(1-\rho-\eta- R)^2}{4 \rho \eta} \right] \!,
\nonumber
\label{eq:C0-AE}
\end{eqnarray}

\begin{eqnarray}
C_{\mu_G^2}^\perp & = &
\frac{R}{2} \left[ - 3 + 5 \rho - 19 \rho^2 + 5 \rho^3 
+ (5 + 28 \rho - 35 \rho^2) \eta - (19 + 35 \rho) \eta^2 + 5 \eta^3 \right] \\
& - & 
6 \left[\rho^2\,
  \ln \frac{(1 + \rho - \eta - R)^2}{4\rho} 
- \eta^2\,\text{ln}\frac{(1+\eta-\rho + R)^2}{4\eta}
- 5 \rho^2 \eta^2 \, \ln \frac{(1-\rho-\eta- R)^2}{4 \rho \eta} \right]\! ,
\nonumber
\label{eq:CmuGsq-AE}
\end{eqnarray}

In addition, we have
\begin{equation}
C_{\mu_\pi^2}^\perp  =  -\frac{C_0}{2} \ , \quad\quad\quad C_{\rhols}^\perp = - C_{\muG}^\perp
\end{equation}

\begin{eqnarray}
C_{\rho_D^3}^\perp & = & 
\frac{R}{6}\Big\{ 77+ 5 \rho^3 + \rho^2 (13-35 \eta) + 13 \eta - 
 59 \eta^2 +  5 \eta^3 - \rho (11 + 12 \eta + 35 \eta^2 )  \Big\}\\
 & + & 
  \, \Bigg\{\eta^2 (10 \rho^2 + 8\eta - 2)
 \, \ln \! \left[ \cfrac{(1-\rho-\eta - R)^2}{4\eta\rho} \right]
 +(8 + 6\rho^2 - 8 \eta - 6\eta^2) \, \ln \! \left[ \cfrac{(1+\rho-\eta-R)^2}{4\rho} \right] \Bigg\},
 \nonumber
\end{eqnarray}

where $R= \sqrt{\rho^2 + (-1+\eta)^2 -2\rho(1+\eta)}$.

\bibliographystyle{jhep} 
\bibliography{main.bib}

\providecommand{\href}[2]{#2}\begingroup\raggedright\begin{thebibliography}{10}

\bibitem{Bordone:2021oof}
M.~Bordone, B.~Capdevila and P.~Gambino, \emph{{Three loop calculations and
  inclusive Vcb}},
  \href{https://doi.org/10.1016/j.physletb.2021.136679}{\emph{Phys. Lett. B}
  {\bfseries 822} (2021) 136679}
  [\href{https://arxiv.org/abs/2107.00604}{{\ttfamily 2107.00604}}].

\bibitem{Bernlochner:2022ucr}
F.~Bernlochner, M.~Fael, K.~Olschewsky, E.~Persson, R.~van Tonder, K.~K. Vos
  et~al., \emph{{First extraction of inclusive $V_{cb}$ from $q^2$ moments}},
  \href{https://arxiv.org/abs/2205.10274}{{\ttfamily 2205.10274}}.

\bibitem{Belle:2021idw}
{\scshape Belle} collaboration, \emph{{Measurements of $q^2$ Moments of
  Inclusive $B \rightarrow X_c \ell^+ \nu_{\ell}$ Decays with Hadronic
  Tagging}}, \href{https://doi.org/10.1103/PhysRevD.104.112011}{\emph{Phys.
  Rev. D} {\bfseries 104} (2021) 112011}
  [\href{https://arxiv.org/abs/2109.01685}{{\ttfamily 2109.01685}}].

\bibitem{Beneke:2020vnb}
M.~Beneke, P.~B\"oer, J.-N. Toelstede and K.~K. Vos, \emph{{QED factorization
  of non-leptonic $B$ decays}},
  \href{https://doi.org/10.1007/JHEP11(2020)081}{\emph{JHEP} {\bfseries 11}
  (2020) 081} [\href{https://arxiv.org/abs/2008.10615}{{\ttfamily
  2008.10615}}].

\bibitem{Beneke:2021jhp}
M.~Beneke, P.~B\"oer, G.~Finauri and K.~K. Vos, \emph{{QED factorization of
  two-body non-leptonic and semi-leptonic B to charm decays}},
  \href{https://doi.org/10.1007/JHEP10(2021)223}{\emph{JHEP} {\bfseries 10}
  (2021) 223} [\href{https://arxiv.org/abs/2107.03819}{{\ttfamily
  2107.03819}}].

\bibitem{Papucci:2021ztr}
M.~Papucci, T.~Trickle and M.~B. Wise, \emph{{Radiative semileptonic $
  \overline{B} $ decays}},
  \href{https://doi.org/10.1007/JHEP02(2022)043}{\emph{JHEP} {\bfseries 02}
  (2022) 043} [\href{https://arxiv.org/abs/2110.13154}{{\ttfamily
  2110.13154}}].

\bibitem{Cali:2019nwp}
S.~Cal\'\i{}, S.~Klaver, M.~Rotondo and B.~Sciascia, \emph{{Impacts of
  radiative corrections on measurements of lepton flavour universality in $B
  \to D \ell \nu_{\ell}$ decays}},
  \href{https://doi.org/10.1140/epjc/s10052-019-7254-x}{\emph{Eur. Phys. J. C}
  {\bfseries 79} (2019) 744}
  [\href{https://arxiv.org/abs/1905.02702}{{\ttfamily 1905.02702}}].

\bibitem{pdg}
{\scshape Particle Data Group} collaboration{\emph{Prog. Theor. Exp. Phys. {\bf
  2020}} {\bfseries 083C01} (2020) }.

\bibitem{Hasenbusch:2018pxj}
J.~Hasenbusch, \emph{{Analysis of inclusive semileptonic $B$ meson decays with
  $\tau$ lepton final states at the Belle experiment}}, Ph.D. thesis, U. Bonn
  (main), 2018.

\bibitem{Mannel:2017jfk}
T.~Mannel, A.~V. Rusov and F.~Shahriaran, \emph{{Inclusive semitauonic $B$
  decays to order ${\mathcal O} (\Lambda_{QCD}^3/m_b^3)$}},
  \href{https://doi.org/10.1016/j.nuclphysb.2017.05.016}{\emph{Nucl. Phys. B}
  {\bfseries 921} (2017) 211}
  [\href{https://arxiv.org/abs/1702.01089}{{\ttfamily 1702.01089}}].

\bibitem{Bernlochner:2022ywh}
F.~U. Bernlochner, Z.~Ligeti, M.~Papucci, M.~T. Prim, D.~J. Robinson and
  C.~Xiong, \emph{{Constrained second-order power corrections in HQET:
  $R(D^{(*)})$, $|V_{cb}|$, and new physics}},
  \href{https://arxiv.org/abs/2206.11281}{{\ttfamily 2206.11281}}.

\bibitem{Mannel:2021mwe}
T.~Mannel, M.~Rahimi and K.~K. Vos, \emph{{Impact of background effects on the
  inclusive $V_{cb}$ determination}},
  \href{https://arxiv.org/abs/2105.02163}{{\ttfamily 2105.02163}}.

\bibitem{Ligeti:2014kia}
Z.~Ligeti and F.~J. Tackmann, \emph{{Precise predictions for $B \to X_c \tau
  \bar \nu$ decay distributions}},
  \href{https://doi.org/10.1103/PhysRevD.90.034021}{\emph{Phys. Rev. D}
  {\bfseries 90} (2014) 034021}
  [\href{https://arxiv.org/abs/1406.7013}{{\ttfamily 1406.7013}}].

\bibitem{Sirlin:1974ni}
A.~Sirlin, \emph{{Radiative corrections to g(v)/g(mu) in simple extensions of
  the su(2) x u(1) gauge model}},
  \href{https://doi.org/10.1016/0550-3213(74)90254-5}{\emph{Nucl. Phys. B}
  {\bfseries 71} (1974) 29}.

\bibitem{Mannel:2010wj}
T.~Mannel, S.~Turczyk and N.~Uraltsev, \emph{{Higher Order Power Corrections in
  Inclusive B Decays}},
  \href{https://doi.org/10.1007/JHEP11(2010)109}{\emph{JHEP} {\bfseries 11}
  (2010) 109} [\href{https://arxiv.org/abs/1009.4622}{{\ttfamily 1009.4622}}].

\bibitem{Mannel:2018mqv}
T.~Mannel and K.~K. Vos, \emph{{Reparametrization Invariance and Partial
  Re-Summations of the Heavy Quark Expansion}},
  \href{https://doi.org/10.1007/JHEP06(2018)115}{\emph{JHEP} {\bfseries 06}
  (2018) 115} [\href{https://arxiv.org/abs/1802.09409}{{\ttfamily
  1802.09409}}].

\bibitem{Fael:2018vsp}
M.~Fael, T.~Mannel and K.~Keri~Vos, \emph{{$V_{cb}$ determination from
  inclusive $b \to c$ decays: an alternative method}},
  \href{https://doi.org/10.1007/JHEP02(2019)177}{\emph{JHEP} {\bfseries 02}
  (2019) 177} [\href{https://arxiv.org/abs/1812.07472}{{\ttfamily
  1812.07472}}].

\bibitem{Gremm:1996df}
M.~Gremm and A.~Kapustin, \emph{{Order 1/m(b)**3 corrections to B $\to$ X(c)
  lepton anti-neutrino decay and their implication for the measurement of
  Lambda-bar and lambda(1)}},
  \href{https://doi.org/10.1103/PhysRevD.55.6924}{\emph{Phys. Rev. D}
  {\bfseries 55} (1997) 6924}
  [\href{https://arxiv.org/abs/hep-ph/9603448}{{\ttfamily hep-ph/9603448}}].

\bibitem{Dassinger:2006md}
B.~M. Dassinger, T.~Mannel and S.~Turczyk, \emph{{Inclusive semi-leptonic B
  decays to order 1 / m(b)**4}},
  \href{https://doi.org/10.1088/1126-6708/2007/03/087}{\emph{JHEP} {\bfseries
  03} (2007) 087} [\href{https://arxiv.org/abs/hep-ph/0611168}{{\ttfamily
  hep-ph/0611168}}].

\bibitem{Falk:1994gw}
A.~F. Falk, Z.~Ligeti, M.~Neubert and Y.~Nir, \emph{{Heavy quark expansion for
  the inclusive decay anti-B ---\ensuremath{>} tau anti-neutrino X}},
  \href{https://doi.org/10.1016/0370-2693(94)91206-8}{\emph{Phys. Lett. B}
  {\bfseries 326} (1994) 145}
  [\href{https://arxiv.org/abs/hep-ph/9401226}{{\ttfamily hep-ph/9401226}}].

\bibitem{Balk:1993sz}
S.~Balk, J.~G. Korner, D.~Pirjol and K.~Schilcher, \emph{{Inclusive
  semileptonic B decays in QCD including lepton mass effects}},
  \href{https://doi.org/10.1007/BF01557233}{\emph{Z. Phys. C} {\bfseries 64}
  (1994) 37} [\href{https://arxiv.org/abs/hep-ph/9312220}{{\ttfamily
  hep-ph/9312220}}].

\bibitem{King:2021xqp}
D.~King, A.~Lenz, M.~L. Piscopo, T.~Rauh, A.~V. Rusov and C.~Vlahos,
  \emph{{Revisiting Inclusive Decay Widths of Charmed Mesons}},
  \href{https://arxiv.org/abs/2109.13219}{{\ttfamily 2109.13219}}.

\bibitem{Czarnecki:1994bn}
A.~Czarnecki, M.~Jezabek and J.~H. Kuhn, \emph{{Radiative corrections to b
  $\to$ c tau anti-tau-neutrino}},
  \href{https://doi.org/10.1016/0370-2693(94)01681-2}{\emph{Phys. Lett. B}
  {\bfseries 346} (1995) 335}
  [\href{https://arxiv.org/abs/hep-ph/9411282}{{\ttfamily hep-ph/9411282}}].

\bibitem{Jezabek:1996db}
M.~Jezabek and L.~Motyka, \emph{{Tau lepton distributions in semileptonic B
  decays}}, \href{https://doi.org/10.1016/S0550-3213(97)00341-6}{\emph{Nucl.
  Phys. B} {\bfseries 501} (1997) 207}
  [\href{https://arxiv.org/abs/hep-ph/9701358}{{\ttfamily hep-ph/9701358}}].

\bibitem{Biswas:2009rb}
S.~Biswas and K.~Melnikov, \emph{{Second order QCD corrections to inclusive
  semileptonic b ---\ensuremath{>} X(c) l anti-nu(l) decays with massless and
  massive lepton}}, \href{https://doi.org/10.1007/JHEP02(2010)089}{\emph{JHEP}
  {\bfseries 02} (2010) 089} [\href{https://arxiv.org/abs/0911.4142}{{\ttfamily
  0911.4142}}].

\bibitem{Fael:2022frj}
M.~Fael, K.~Sch\"onwald and M.~Steinhauser, \emph{{A first glance to the
  kinematic moments of $B \to X_c \ell \nu$ at third order}},
  \href{https://arxiv.org/abs/2205.03410}{{\ttfamily 2205.03410}}.

\bibitem{Gambino:2004qm}
P.~Gambino and N.~Uraltsev, \emph{{Moments of semileptonic B decay
  distributions in the 1/m(b) expansion}},
  \href{https://doi.org/10.1140/epjc/s2004-01671-2}{\emph{Eur. Phys. J. C}
  {\bfseries 34} (2004) 181}
  [\href{https://arxiv.org/abs/hep-ph/0401063}{{\ttfamily hep-ph/0401063}}].

\bibitem{Gambino:2007rp}
P.~Gambino, P.~Giordano, G.~Ossola and N.~Uraltsev, \emph{{Inclusive
  semileptonic B decays and the determination of |V(ub)|}},
  \href{https://doi.org/10.1088/1126-6708/2007/10/058}{\emph{JHEP} {\bfseries
  10} (2007) 058} [\href{https://arxiv.org/abs/0707.2493}{{\ttfamily
  0707.2493}}].

\bibitem{Fael:2020iea}
M.~Fael, K.~Sch\"onwald and M.~Steinhauser, \emph{{Kinetic Heavy Quark Mass to
  Three Loops}},
  \href{https://doi.org/10.1103/PhysRevLett.125.052003}{\emph{Phys. Rev. Lett.}
  {\bfseries 125} (2020) 052003}
  [\href{https://arxiv.org/abs/2005.06487}{{\ttfamily 2005.06487}}].

\bibitem{Freytsis:2015qca}
M.~Freytsis, Z.~Ligeti and J.~T. Ruderman, \emph{{Flavor models for $\bar{B}
  \to D^{(*)} \tau \bar{\nu}$}},
  \href{https://doi.org/10.1103/PhysRevD.92.054018}{\emph{Phys. Rev. D}
  {\bfseries 92} (2015) 054018}
  [\href{https://arxiv.org/abs/1506.08896}{{\ttfamily 1506.08896}}].

\bibitem{Fael:2019umf}
M.~Fael, T.~Mannel and K.~K. Vos, \emph{{The Heavy Quark Expansion for
  Inclusive Semileptonic Charm Decays Revisited}},
  \href{https://doi.org/10.1007/JHEP12(2019)067}{\emph{JHEP} {\bfseries 12}
  (2019) 067} [\href{https://arxiv.org/abs/1910.05234}{{\ttfamily
  1910.05234}}].

\bibitem{Gambino:2010jz}
P.~Gambino and J.~F. Kamenik, \emph{{Lepton energy moments in semileptonic
  charm decays}},
  \href{https://doi.org/10.1016/j.nuclphysb.2010.07.019}{\emph{Nucl. Phys. B}
  {\bfseries 840} (2010) 424}
  [\href{https://arxiv.org/abs/1004.0114}{{\ttfamily 1004.0114}}].

\bibitem{CLEO:2009uah}
{\scshape CLEO} collaboration, \emph{{Measurement of absolute branching
  fractions of inclusive semileptonic decays of charm and charmed-strange
  mesons}}, \href{https://doi.org/10.1103/PhysRevD.81.052007}{\emph{Phys. Rev.
  D} {\bfseries 81} (2010) 052007}
  [\href{https://arxiv.org/abs/0912.4232}{{\ttfamily 0912.4232}}].

\bibitem{WorkInProgress3}
M.~Fael and K.~K. Vos, \emph{{Work in Progress}}, .

\bibitem{Leljak:2021vte}
D.~Leljak, B.~Meli\'c and D.~van Dyk, \emph{{The $ \overline{B} $
  \textrightarrow{} \ensuremath{\pi} form factors from QCD and their impact on
  |V$_{ub}$|}}, \href{https://doi.org/10.1007/JHEP07(2021)036}{\emph{JHEP}
  {\bfseries 07} (2021) 036}
  [\href{https://arxiv.org/abs/2102.07233}{{\ttfamily 2102.07233}}].

\bibitem{Belle:2021eni}
{\scshape Belle} collaboration, \emph{{Measurements of Partial Branching
  Fractions of Inclusive $B \to X_u \, \ell^+\, \nu_{\ell}$ Decays with
  Hadronic Tagging}},
  \href{https://doi.org/10.1103/PhysRevD.104.012008}{\emph{Phys. Rev. D}
  {\bfseries 104} (2021) 012008}
  [\href{https://arxiv.org/abs/2102.00020}{{\ttfamily 2102.00020}}].

\bibitem{Bernlochner:2021vlv}
F.~U. Bernlochner, M.~F. Sevilla, D.~J. Robinson and G.~Wormser,
  \emph{{Semitauonic b-hadron decays: A lepton flavor universality
  laboratory}}, \href{https://doi.org/10.1103/RevModPhys.94.015003}{\emph{Rev.
  Mod. Phys.} {\bfseries 94} (2022) 015003}
  [\href{https://arxiv.org/abs/2101.08326}{{\ttfamily 2101.08326}}].

\bibitem{Bernlochner:2017jka}
F.~U. Bernlochner, Z.~Ligeti, M.~Papucci and D.~J. Robinson, \emph{{Combined
  analysis of semileptonic $B$ decays to $D$ and $D^*$: $R(D^{(*)})$,
  $|V_{cb}|$, and new physics}},
  \href{https://doi.org/10.1103/PhysRevD.95.115008}{\emph{Phys. Rev. D}
  {\bfseries 95} (2017) 115008}
  [\href{https://arxiv.org/abs/1703.05330}{{\ttfamily 1703.05330}}].

\bibitem{WorkInProgress}
T.~Mannel, M.~Rahimi and K.~K. Vos, \emph{{Work in Progress}}, .

\end{thebibliography}\endgroup

\end{document}